\documentclass[11pt]{article} \textwidth 170mm \textheight 225mm
\usepackage{amsfonts}
\usepackage{amssymb}
\usepackage{amsthm}
\usepackage{amsmath}
\usepackage{newlfont}
\usepackage{newlfont}
\usepackage{bm}
\usepackage{txfonts}
\usepackage{graphics}
\usepackage{graphicx}
\usepackage{color}
\usepackage{indentfirst}

\begin{document}
\title{ The third-order elastic moduli and pressure derivatives for AlRE (RE=Y, Pr, Nd, Tb, Dy, Ce) intermetallics with B2-structure: A first-principles study  \footnote{The work is supported by the National Natural Science Foundation
of China (10774196) and Project No. CDJXS11102211 supported by the Fundamental
Research Funds for the Central Universities. }}
\author{Rui Wang\footnote{rcwang@cqu.edu.cn; Tel: +86 13527528737}, Shaofeng Wang, and Xiaozhi Wu\\
{ \small Institute for Structure and Function and Department of
Physics, Chongqing University, }\\{\small Chongqing 400044, People's
Republic of China.}}
\date{}
\maketitle

\begin{abstract}
\baselineskip 15pt
The third-order elastic moduli and pressure derivatives of the second-order elastic constants of novel B2-type AlRE (RE=Y, Pr, Nd, Tb, Dy, Ce) intermetallics  are presented from first-principles  calculations. The elastic moduli are obtained from the coefficients of the polynomials from the nonlinear least-squares fitting of the energy-strain functions. The calculated second-order elastic constants of AlRE intermetallics are consistent
with the previous calculations. To judge that our computational accuracy is reasonable, the calculated
third-order constants of Al are compared with the available experimental data and other theoretical results
and found very good agreement. In comparison with the theory of the linear elasticity, the third-order effects are very important with the finite strains are lager than approximately 3.5\%. Finally, the pressure derivative has been discussed.

\end{abstract}
\vskip 0.1in{\small  } \vskip 0.2in

\vskip 0.1in

 \vskip 0.3in

\baselineskip 20pt

\section{Introduction}

Aluminium alloys are widely used in the automotive, aircraft and aerospace industries and interest in their applications is increasing. Rare-earth element is a class of special elements which are known to be the most effective modifier and recrystallizator to the Al alloys. Recently, the Al-TM-RE (TM is a transition metal, and RE denotes a rare-earth element) systems have received great attention due to their unique mechanical properties, such as high tensile strength, good ductility, high corrosion resistance and thermal stability \cite{Rizzi,Antonowicz,Miracle,Kelton,Yang,Shpak}. Recently, a new class of highly ordered and ductile intermetallic compounds with the CsCl-type B2 structure in space group Pm3m (such as YAg and YCu) has been discovered with guidance from Gschneider \cite{Gschneidner} and Morris et al.\cite{Morris}. It is known that the B2 structure is a very important intermetallics, so the B2-AlRE alloys as well as B2-MgRE alloys have been investigated intensely \cite{Tao,Tao1,Tao2,Srivastava,Asta,Ugur,Wu,Hu}. Some experimental information of structures for Al-RE equiatomic compounds can be found in literature\cite{Villars}, and for example, the Al-Y, Al-Ce, Al-Nd, Al-Dy, Al-Pr, etc. are available for metastable B2 phase \cite{Dagerhamn,Yakunin,Stillwell,Baenziger,Kripjakevich}. More recently, various studies have been undertaken of the bulk, elastic, thermodynamic, and electronic properties for the rare-earth-aluminium B2-AlRE intermetallics from first-principles calculations \cite{Tao,Tao1,Tao2,Srivastava,Asta}, and these results are very important to the novel material design and further scientific and technical investigations.

For technical applications, applying finite deformation to materials
is also very important due to the  AlRE intermetallics with superior
mechanical properties, while the theory of linear elasticity with assuming infinitesimal strain isn't proper any longer and instead the theory of nonlinear elasticity is required \cite{Birch,Murnaghan,Second,Brugger,Thurston}. In the linear
theory of elasticity, the second-order elastic constants (SOECs) are sufficient to
describe the elastic stress-strain response and wave propagation in
solids\cite{Born}. In nonlinear elastic theory, high-order elastic
constants, such as third-order elastic constants (TOECs),  play an
important role as well as SOECs \cite{Thurston}. TOECs are not only
used in describing mechanical phenomena when applying large stress
or strain, but also can be used to describe the anharmonic
properties such as thermal expansion, the interaction of acoustic
and thermal phonon, changes in acoustic velocities due to elastic
strain, etc.\cite{Wallace,Hiki, Thurston1}. In addition, the elastic properties related high pressure can also be discussed using the SOECs and TOECs directly.  Though many experiments have been performed to determined SOCEs and high-order elastic constants \cite{Second}, to obtain a complete set of TOECs is still not a simple task, especially for crystals with low symmetry or with low-yield stress. Recently, a simple
method using first-principles calculations have been
 employed to determined TOECs \cite{Michal1,Zhao,Wang} as well as high-order elastic constants, and their results show
good agreement with experiments. In this approach, the homogeneous deformation strain is applied to the system and usually simple deformation modes are used such as uniaxial tension or compression, simple or pure shear, and other combinations of homogenous strains. To provide significant information with respect to application and design of these novel AlRE alloys, to study further the nonlinear elastic constants such as TOECs of these compounds is required.

\section{Computational methods}

Our method for calculating elastic constants mentions finite-strain
continuum elasticity theory. For a solid-body with a finite deformation, the configuration of a material point in the system after deformation is represented as $x'=x'(x)$, where $x$ is the initial configuration at the equilibrium state. The relation between the strained and unstrained crystal is constructed by the deformation tensor
\begin{equation}
J_{ij}=\frac{\partial x'_{i}}{\partial
x_{j}}
\end{equation}
where $i$ and $j$(=1,2,3)represent Cartesian coordinates. The relation between the elastic constants and the strain energy density, $\Phi$, can
be written as
\begin{equation}\label{Phi}
\Phi(\bm{\eta})
=\frac{1}{2!}\sum_{ijkl}C_{ijkl}\eta_{ij}\eta_{kl}+\frac{1}{3!}\sum_{ijklmn}C_{ijklmn}\eta_{ij}\eta_{kl}\eta_{mn}+\ldots,
\end{equation}
where $\eta_{ij}$ is Lagrangian strain tensor, which
is defined as \cite{Thurston}
\begin{equation}\label{eta}
\eta_{ij}=\frac{1}{2}\bigg(\sum_{k}J_{ik}J_{jk}-\delta_{ij}\bigg).
\end{equation} The n$th$-order ($n\geq2$) elastic constants was defined by
Brugger \cite{Brugger} as
\begin{equation}\label{s2}
C_{ijklmn\ldots}=C_{IJK\ldots}=\frac{{\partial}^{n}\Phi}{\partial
\eta_{ij}
\partial \eta_{kl} \partial \eta_{mn}\cdots}\bigg|{_{\eta=0}},
\end{equation}
where $I$, $J$, and $K$ are Voigt subscripts; and the Lagrangian
strain tensor $\bm\eta$ links this notation by
\begin{equation}\label{etas}
{\bm\eta}=(\eta_{1}, \eta_{2}, \eta_{3}, \eta_{4}, \eta_{5}, \eta_{6}).
\end{equation}
Because of B2-AlRE intermetallics with cubic
symmetry, there are three independent SOECs ($C_{11}$, $C_{12}$,
$C_{44}$) and six TOECs ($C_{111}$, $C_{112}$, $C_{123}$,$C_{144}$,
$C_{155}$, $C_{456}$). To calculate the complete SOECs and TOECs, we introduced six Lagrangian strain tensors in terms of a single parameter $\xi$. Inserting these strains into Eq. (\ref{Phi}), the elastic energy per unit mass can be written as an expansion in the strain parameter $\xi$,
\begin{equation}\label{ex}
\Phi(\bm{\eta})
=\frac{1}{2}\Lambda_{2}\xi^2+\frac{1}{6}\Lambda_{3}\xi^3+O(\xi^4),
\end{equation}
where the coefficients $\Lambda_{2}$ and $\Lambda_{3}$ are combinations of second- and third-order elastic constants of the crystal, respectively. The selections of the different deformation modes leads to different strains used in this work, which are labeled as $\bm{\eta}_{\alpha}$, $\alpha=A,B,\ldots,F$. The applied strains and the corresponding coefficients $\Lambda_{2}$ and $\Lambda_{3}$ are listed in Table 1.

In every case for $\bm\eta_{\alpha}$, $\xi$ is varied between -0.08
and 0.08 with step 0.008 to obtain accurate TOECs. The maximal amplitude of the deformations is reasonable to obtained the TOECs accurately \cite{Michal1,Zhao}.  The elastic
constants will be obtained from the least-square polynomial fit to
the strain-energy relation from first-principles total-energy
calculations. In order to obtain the unit cell of strained crystal,
the deformation tensor $J_{ij}$ is applied to the undeformed
 lattice vectors $\textbf{a}_{i}$, where $i$ is the lattice index.
The deformed crystal is obtained then from
$\mathbf{a}_{i}'=\sum_{j}J_{ij}\mathbf{a}_{j}$. To implement the
different deformation modes in our calculation, we need to have the
deformation tensor $J_{ij}$, which is determined from the Lagrangian
strain by inverting Eq. (\ref{eta}),
\begin{equation}\label{J}
J_{ij}=\delta_{ij}+\eta_{ij}-\frac{1}{2}\sum_{k}\eta_{ik}\eta_{kj}+
\frac{1}{2}\sum_{kl}\eta_{ik}\eta_{kl}\eta_{lj}+\cdots
\end{equation}
 For a given $\eta$, in general, $J$ is not unique but this is not a
problem since the Lagrange strain brings rotational invariance.

We carry out first-principles total-energy calculations based on the
density functional theory (DFT) level, using the ¡±Vienna ab initio
simulation package¡± (VASP 4.6) developed at the Institut f\"{u}r
Materialphysik of Universit\"{a}t Wien \cite{Kresse1, Kresse2,
Kresse3}. The Perdew-Burke-Ernzerhof (PBE) \cite{Perdew1,Perdew2}
exchange-correlation functional for the
generalized-gradient-approximation(GGA) is used. A plane-wave basis
set is employed within the framework of the projector augmented wave
(PAW) method \cite{Blochl, Kresse4}.  The
equilibrium theoretical crystal structures are determined by
minimizing the Hellmann-Feynman force on the atoms and the stress on
the unit cell. The convergence of energy and force are set to
$1.0\times10^{-6}eV$ and $1.0\times10^{-4}eV/{\AA}$, respectively.  For the Billouin zone
integrals, reciprocal space is represent by Monkhorst-Pack special
k-point scheme \cite{Monkhorst}. Since high accuracy is needed to evaluate the TOECs, the $k-$point mesh size with $25\times25\times25$ and cutoff
energy with 600eV are used to calculate SOECs and TOECs for all calculated materials. Our convergence tests show that the chosen parameters are sufficient to reach the desired convergence for the total energy as well as the TOECs. In Table 2, we give the equilibrium
lattice constants for AlRE intermetallics in our calculation,
and it is shown that the results agree well with the previous
calculation \cite{Tao1,Srivastava} and experiment \cite{Dagerhamn,Yakunin,Stillwell,Baenziger,Kripjakevich}.

\section{ Results and discussion}

 For completeness, we also list our calculated SOECs as well as the previous results by Tao and Ouyang, et al.\cite{Tao1} for AlRE intermetallics in Table 2. Our results show good agreement with those obtaining from previous first-principles calculations. The bulk modulus $B=(C_{11}+2C_{12})/3$ in our calculations also provide extremely good agreement with the values fitted from the Rose's equation of state \cite{Tao1}. To benchmark
the reliability results of the presented method, we have compared
our theoretical results for Al, which is a
well-studied cubic crystal having $fcc$ structure with available experimental data \cite{Second,Rose} and previous theoretical findings
within DFT theory \cite{Wang}. The calculated SOECs and TOECs of Al are listed in Table 3. In the previous calculations, the all third-order
elastic constants of Al were obtained \cite{Wang}.  Overall, our ab initio
results agree well with the previous calculations and the
experimental data for both SOECs and TOECs. We present the unknown values of TOECs for all calculated AlRE intermetallics in Table 4.
The six strain-energies for AlY,
including the results of the first principles calculations and the
fitted polynomials, are shown Figure \ref{figure1}. It shows that the
strain-energies with negative strains are always larger than those
with positive strains, so the values of TOECs are typically
negative. It is worth noting that for the studied intermetallics
with Lagrangian strains up to 8.0$\%$, including the terms up to
third-order in energy expansion sufficed to obtain good agreement
with our ab initio results. Next we focus on examining for which range of strains the
third-order effects dominant the properties of solids. In Figure \ref{figure2}, we
also show the curves of the linear elasticity comparison with the
nonlinear elasticity and the DFT results particular deformation
$\bm{\eta}_{A}$ in AlY. It is clearly to see that the third-order
effects must be considered and it is not sufficient for linear
elasticity when applied to finite deformations larger than
approximately $3.5\%$, while the nonlinear elasticity must be considered.

\section{The effective elastic constants under pressure}
 In the case of materials under larger hydrostatic
pressure, it is useful to describe the  nonlinear elastic properties
using the concept of effective elastic constants $C_{ij}(P)$.
Usually, $C_{ij}(P)$ as a function of pressure $P$ can be expanded
as Taylor series and it is appropriate to consider only linear term
in the external hydrostatic pressure
\begin{equation}
  C_{IJ}(P) \approx C_{IJ}+\frac{dC_{IJ}(P)}{dP}P= C_{IJ}+C_{ij}'P,
\end{equation}
with pressure derivatives $C_{IJ}' $  being material parameters. Once the complete SOECs and TOECs  are available, the the effective elastic constants under high pressure can be discussed directly.
Naturally, the values of $C_{IJ}'$ are determined from components of
both SOECs and TOECs, and can be expressed as \cite{Birch}
\begin{eqnarray}
&&C_{11}'=-\frac{C_{111}+2C_{112}+2C_{11}+2C_{12}}{C_{11}+2C_{12}} \nonumber ,\\
&&C_{12}'=-\frac{2C_{112}+C_{123}-C_{11}-C_{12}}{C_{11}+2C_{12}} \nonumber ,\\
&&C_{44}'=-\frac{2C_{155}+C_{144}+C_{11}+2C_{12}+C_{44}}{C_{11}+2C_{12}},
\end{eqnarray}
while the pressure derivative of the bulk modulus is expressed as
\begin{equation}
B'=\frac{C_{11}'+2C_{12}'}{3}.
\end{equation}

We give the values of pressure derivatives $C_{IJ}'$ and $B'$
on the basis of our prediction for second- and third-order elastic
constants in Table 5. We also list the results of $B'$ obtained from the Rose's equation of state \cite{Tao1}, which show good agreement with the predicted results from TOECs in our calculations. In Ref.\cite{Majewski}, the following method
has been used to determine of the pressure dependence of the
second-order elastic constants. First, the hydrostatic strain has
been applied to the crystal, and then the crystal has been
additionally deformed to determine the pressure dependent elastic
constants. The ab initio results for the total elastic energy
combined with the strain-energy relation have enabled us to
determine $C_{IJ}(P)$ and $C_{IJ}'$. The methods of calculations on
the pressure derivatives had also been employed in
Ref.\cite{Michal1} in which the related properties have been studied for
selected semiconductors. Therefore, we believe that the method used
in our paper is accurate. Furthermore, the experiments when
materials are under high pressure may obtain pressure derivatives,
and can demonstrate that our results of the nonlinear elasticity for
AlRE intermetallics are reliable.

\section{Conclusions}

In this work, we present the unknown TOECs $C_{IJK}$ and the pressure derivatives $C_{IJ}'$ of the SOCEs for the novel AlRE (RE=Y, Pr, Nd, Tb, Dy, Ce) intermetallics with B2-structure using the first-principles total-energy calculations combined with the method of applying a series of homogenous finite strains to the crystals. The PAW method within the GGA is employed. From the nonlinear least-squares fitting, the predictions for the SOECs and TOECs are obtained from the coefficients of the fitted polynomials of the energy-strain functions. The calculated second-order elastic constants of AlRE intermetallics are consistent with the previous calculations.  To benchmark
the reliability results of the presented method, we have computed the TOECs for Al and compared with the available experimental data and other theoretical results and found very good agreement.  In comparison with the theory of the linear elasticity, our work shows the nonlinear elastic effects must be considered when the finite deformations applied are lager than approximately 3.5\%. It is also worth noting that for the studied intermetallics and examined range of deformations, including the terms up to third-order in energy functions sufficed to obtained good agreement with our calculated results. We have discussed the pressure derivatives of the SOECs within the framework of the second- and third-order elastic constants, and the pressure derivatives of the bulk modulus agree well with those obtained from Rose's equation of state. We believe that our
DFT results of the predicted TOECs can be a very useful
tool in applying these novel rare-earth intermetallics to practical engineering, in which nonlinear effects often really matter.

 \vskip 1in

\footnotesize

\def\refname{{\large\bfseries References}}

\newpage

\noindent{{\bf Table 1} The coefficients $\Lambda_{2}$ and $\Lambda_{3}$ in Eq .(\ref{ex}) of corresponding Lagrangian strains as combinations of SOECs and TOECs for crystals with cubic symmetry.

\begin{tabular}{ccc}
  \hline
  Strain type& $\Lambda_{2}$ & $\Lambda_{3}$ \\
  \hline
  $\bm{\eta}_{A}=(\xi, 0, 0, 0, 0, 0)$ & $\frac{1}{2}C_{11}$ & $\frac{1}{6}C_{111}$ \\
  $\bm{\eta}_{B}=(\xi, \xi, 0, 0, 0, 0)$ & $(C_{11}+C_{12}) $& $\frac{1}{3}C_{111}+C_{112} $\\
  $\bm{\eta}_{C}=(\xi, \xi, \xi, 0, 0, 0)$ &$ \frac{3}{2}C_{11}+3C_{12}$ & $\frac{1}{2}C_{111}+3C_{112}+C_{123}$ \\
  $\bm{\eta}_{D}=(\xi, 0, 0, \xi, 0, 0)$ & $\frac{1}{2}C_{11}+\frac{1}{2}C_{44}$ & $\frac{1}{6}C_{111}+\frac{1}{2}C_{144}$ \\
  $\bm{\eta}_{D}=(\xi, 0, 0, 0, \xi, 0)$ & $\frac{1}{2}C_{11}+\frac{1}{2}C_{44}$ & $\frac{1}{6}C_{111}+\frac{1}{2}C_{155}$ \\
    $\bm{\eta}_{D}=(0, 0, 0, \xi, \xi, \xi)$ &$ \frac{3}{2}C_{44}$ & $C_{456}$ \\
  \hline
\end{tabular}
\vskip0.5in

\noindent{{\bf Table 2} Our calculated lattice constants for AlRE (RE=Y, Pr, Nd, Tb, Dy, Ce) intermetallics in comparison with the values of the previous calculations and the experiments. The second-order elastic constants (SOECs) and bulk modulus are also listed.

\begin{tabular}{ccccccc}
  \hline
   & AlY & AlPr & AlNd & AlTb & AlDy & AlCe \\
  \hline
  $a$ ({\AA}) & 3.606$^{a}$& 3.760$^{a}$& 3.729$^{a}$& 3.614$^{a}$& 3.597$^{a}$& 3.675$^{a}$ \\
  & 3.605$^{b}$, 3.522$^{c}$, 3.754$^{d}$&3.759$^{b}$, 3.533$^{c}$, 3.82$^{e}$, &3.728$^{b}$, 3.75$^{f}$&3.614$^{b}$&3.596$^{b}$, 3.71$^{g}$&3.795$^{b}$, 3.575$^{c}$, 3.85$^{h}$ \\
  $C_{11}$ (GPa)& 81.26$^{a}$, 77.93$^{b}$ & 72.61$^{a}$, 66.83$^{b}$ & 74.25$^{a}$, 69.00$^{b}$ & 80.98$^{a}$, 77.13$^{b}$ & 81.40$^{a}$, 78.28$^{b}$& 62.66$^{a}$, 64.75$^{b}$ \\
  $C_{12}$ (GPa)& 54.58$^{a}$, 56.58$^{b}$ & 46.82$^{a}$, 49.26$^{b}$ & 48.71$^{a}$, 51.09$^{b}$ & 55.17$^{a}$, 56.92$^{b}$ & 56.11$^{a}$, 57.73$^{b}$& 52.64$^{a}$, 47.24$^{b}$ \\
  $C_{44}$ (GPa)& 62.79$^{a}$, 61.91$^{b}$ & 47.30$^{a}$, 46.67$^{b}$ & 50.24$^{a}$, 49.83$^{b}$ & 62.30$^{a}$, 61.17$^{b}$ & 64.00$^{a}$, 63.06$^{b}$& 42.53$^{a}$, 43.31$^{b}$ \\
  B(GPa)        & 63.47$^{a}$, 63.70$^{b}$ & 55.42$^{a}$, 55.11$^{b}$ & 57.22$^{a}$, 57.06$^{b}$ & 63.73$^{a}$, 63.66$^{b}$ & 64.54$^{a}$, 64.58$^{b}$& 55.98$^{a}$, 53.07$^{b}$ \\
  \hline
\end{tabular}

\begin{tabular}{c}
  \leftline {${}^{a}$This work}\\
  \leftline {${}^{b}$Reference\cite{Tao1}}\\
  \leftline {${}^{c}$Reference\cite{Srivastava}}\\
  \leftline {${}^{d}$Reference\cite{Dagerhamn}} \\
  \leftline {${}^{e}$Reference\cite{Kripjakevich}}\\
  \leftline {${}^{f}$Reference\cite{Stillwell}}\\
  \leftline {${}^{g}$Reference\cite{Baenziger}}\\
  \leftline {${}^{h}$Reference\cite{Yakunin}}
\end{tabular}

\vskip 0.5in
\noindent{{\bf Table 3}.  Comparison of the calculated
SOECs and TOECs for Al with the previous theoretical values and
experimental results. The unit of all data is GPa.

\begin{tabular}{cccc}
  \hline
  \ \ & This work & Previous study & Experiment \\
  \hline
  $C_{11}$&113.21& $110.4^{a}$& $108^{b}$  \\
  $C_{12}$&53.93&  $54.5^{a}$&$63.92^{b}$\\
  $C_{44}$&30.9& $31.3^{a}$&$28.3^{b}$\\
  $C_{111}$&-1088.20&  $-1253^{a}$&$-1080^{c}$, $-1224^{c}$,$-1427^{d}$\\
  $C_{112}$&-415.40& $-426^{a}$&$-315^{c}$, $-373^{c}$,$-408^{d}$\\
  $C_{144}$&-14.25& $-12^{a}$&$-23^{c}$, $-64^{c}$,$-85^{d}$\\
  $C_{155}$&-479.58&$-493^{a}$&$-340^{c}$, $-368^{c}$,$-396^{d}$\\
  $C_{123}$&167.52& $153^{a}$&$36^{c}$, $25^{c}$,$32^{d}$\\
  $C_{456}$&-24.18&$-21^{a}$&$-30^{c}$, $-27^{c}$,$-42^{d}$\\
  \hline
\end{tabular}

\begin{tabular}{cccccccc}
  \leftline {${}^{a}$Reference\cite{Wang}} \\
  \leftline {${}^{b}$Reference\cite{Second} }\\
  \leftline {${}^{c}$ At 300K and 298K from References \cite{Second} and \cite{Rose} }\\
  \leftline {${}^{d}$ At 80K from Reference\cite{Second} }\\
\end{tabular}}

\newpage
\noindent{{\bf Table 4}.  The predicted results of the third-order elastic constants (TOECs) of AlRE (RE=Y, Pr, Nd, Tb, Dy, Ce) intermetallics. The unit of all data is GPa.

\begin{tabular}{ccccccc}
  \hline
   & AlY & AlPr & AlNd & AlTb & AlDy & AlCe \\
   \hline
  $C_{111}$ & -830.76 & -662.54 & -705.11 &-839.01 & -830.318 & -275.34 \\
  $C_{112}$ & -170.00 & -136.39 & -137.27 & -168.03 & -170.36 & -160.98 \\
  $C_{123}$ & -200.28 & -202.06 & -231.37 & -199.37 & -210.12 & -264.29 \\
  $C_{144}$ & -183.22 &  -44.48& -53.57 & -154.28 &  -167.32 & 97.32 \\
  $C_{155}$ & -355.51 & -290.12 & -300.70 & -348.44 & -352.47 & -269.64 \\
  $C_{456}$ & -321.66 & -288.12 & -295.93 & -317.28 & -321.05 &  -362.77 \\
  \hline
\end{tabular}

\vskip 0.2in
\noindent{{\bf Table 5}.  Predictions for the pressure derivatives
 of SOECs for AlRE (RE=Y, Pr, Nd, Tb, Dy, Ce) intermetallics. The pressure derivatives of the bulk modulus as well as the values obtained from Rose' equation of state \cite{Tao1} are shown.

\begin{tabular}{ccccccc}
  \hline
   & AlY & AlPr & AlNd & AlTb & AlDy & AlCe \\
   \hline
  $C_{11}'$ & 4.72 & 4.19 & 4.27 & 4.72 & 4.62 & 2.18 \\
  $C_{12}'$ & 3.55 & 3.57 & 3.66 & 3.51 & 3.56 & 4.18 \\
  $C_{44}'$ & 3.37 & 2.47 & 2.52 & 3.13 & 3.17 & 1.38 \\
  $B'$      & 3.96 & 3.87 & 3.89 & 3.92 & 3.93 & 3.57 \\
  $B'$$^{\cite{Tao1}}$& 3.99 & 3.92 & 3.95 &3.92 & 3.99 & 3.72 \\
  \hline
\end{tabular}

 \newpage
\begin{figure}
\scalebox{0.6}[0.6]{\includegraphics{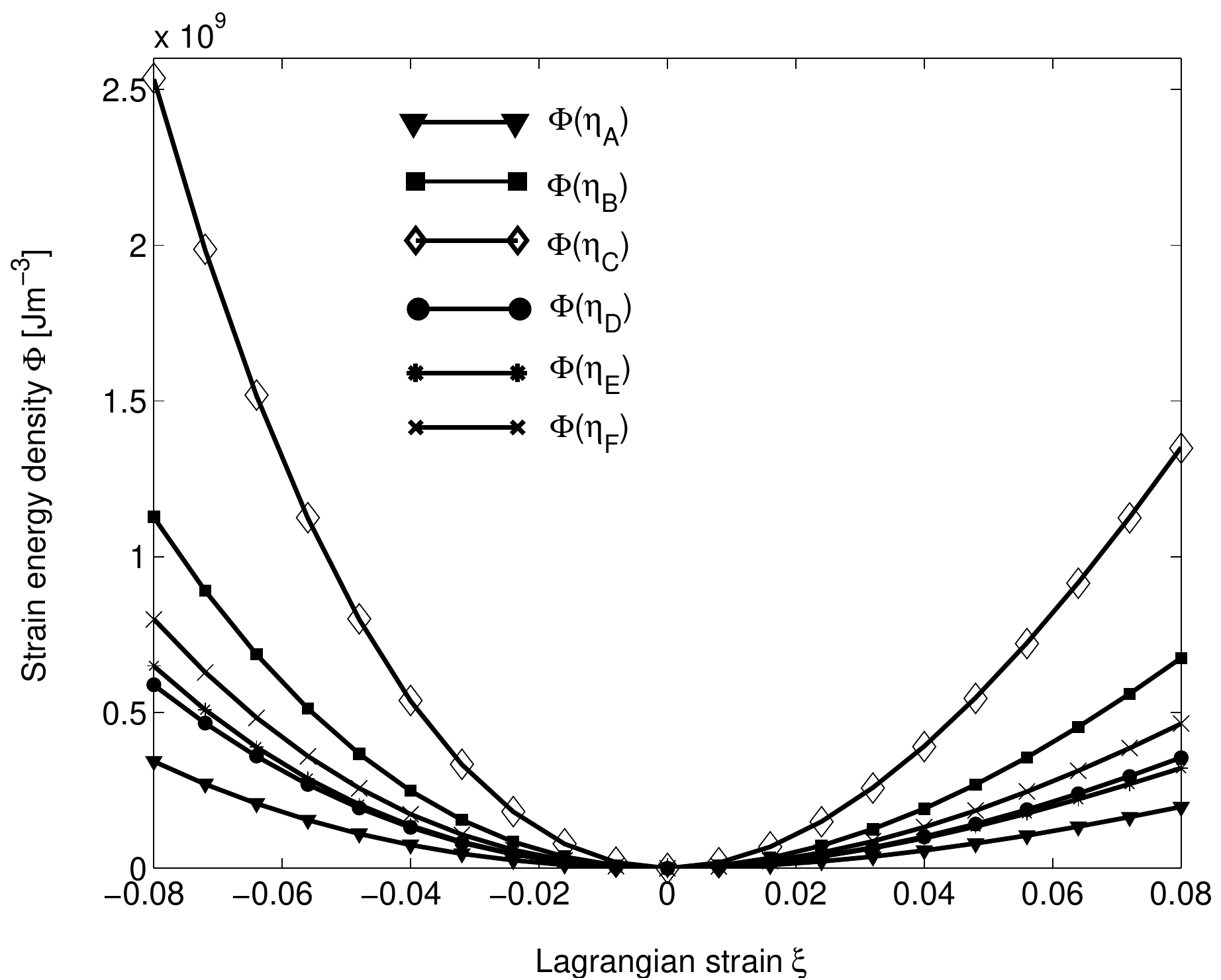}}
\caption{The strain-energy relations for AlY. The discrete points denote the values of DFT calculations; solid curves represent the results obtained from third-order polynomial fitting.}
\label{figure1}
\end{figure}

\begin{figure}
\scalebox{0.6}[0.6]{\includegraphics{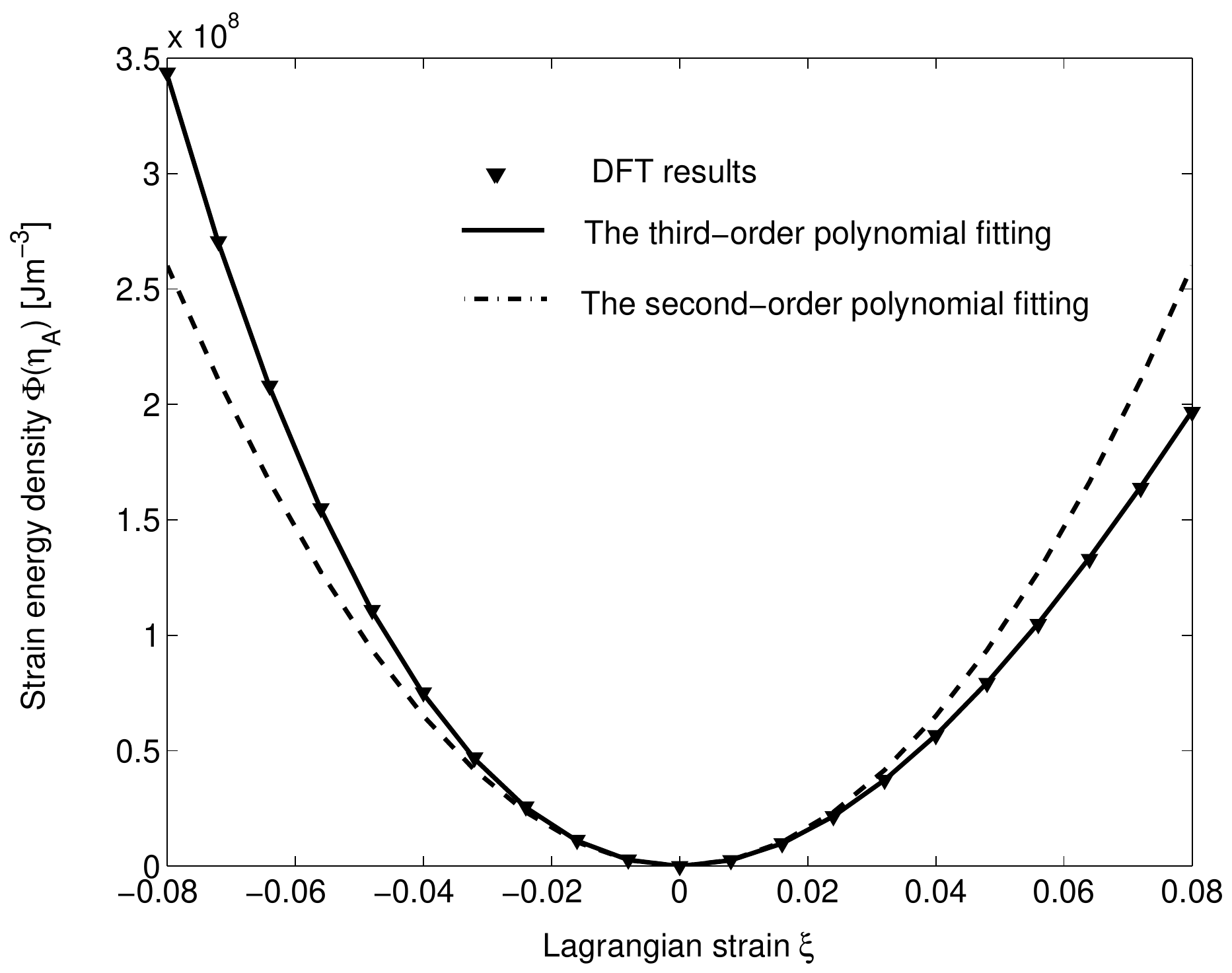}}
\caption{Energy as a function of Lagrangian strain parameter $\xi$ for particular deformation $\bm{\eta}_{A}$for AlY. Full points denote results of DFT computations; solid and dashed curves indicate the results obtained from nonlinear and linear elasticity theory, respectively.}
\label{figure2}
\end{figure}

\end{document}